\documentclass{article}

     \usepackage[nonatbib, final]{neurips_2021}

\usepackage[utf8]{inputenc}       % allow utf-8 input
\usepackage[T1]{fontenc}          % use 8-bit T1 fonts
\usepackage{hyperref}             % hyperlinks
\usepackage{url}                  % simple URL typesetting
\usepackage{booktabs}             % professional-quality tables
\usepackage{amsfonts}             % blackboard math symbols
\usepackage{nicefrac}             % compact symbols for 1/2, etc.
\usepackage{microtype}            % microtypography
\usepackage[dvipsnames]{xcolor}   % colors
\usepackage{multirow}
\usepackage{color,soul}
\usepackage[bottom,hang,flushmargin]{footmisc}

\usepackage{tabularx}
\usepackage{threeparttable}
\usepackage{booktabs}
\usepackage{bm}
\usepackage{amsmath}
\usepackage{stfloats}
\usepackage{gensymb}
\usepackage{graphicx}
\usepackage{tikz}
\usepackage[square,numbers,sort&compress]{natbib}
\title{Anatomical and Diagnostic Bayesian Segmentation in Prostate MRI ---Should Different Clinical Objectives Mandate Different Loss Functions?}

\author{
  \hspace{-3.5mm} Anindo Saha{\tiny $^1$}, Joeran Bosma{\tiny $^1$}, Jasper Linmans{\tiny $^{1,2}$}, Matin Hosseinzadeh{\tiny $^1$}, Henkjan Huisman{\tiny $^1$} \\
  {\tiny $^1$}Diagnostic Image Analysis Group, Radboud University Medical Center, The Netherlands \\
  {\tiny $^2$}Computational Pathology Group, Radboud University Medical Center, The Netherlands \\
  \texttt{\{anindya.shaha,joeran.bosma,jasper.linmans,}         \\
  \texttt{matin.hosseinzadeh,henkjan.huisman\}@radboudumc.nl}   \\
}

\begin{document}

\maketitle

\begin{abstract}
    We hypothesize that probabilistic voxel-level classification of anatomy and malignancy in prostate MRI, although typically posed as near-identical segmentation tasks via U-Nets, require different loss functions for optimal performance due to inherent differences in their clinical objectives. We investigate distribution, region and boundary-based loss functions for both tasks across 200 patient exams from the publicly-available ProstateX dataset. For evaluation, we conduct a thorough comparative analysis of model predictions and calibration, measured with respect to multi-class volume segmentation of the prostate anatomy (whole-gland, transitional zone, peripheral zone), as well as, patient-level diagnosis and lesion-level detection of clinically significant prostate cancer. Notably, we find that distribution-based loss functions (in particular, focal loss) are well-suited for diagnostic or panoptic segmentation tasks such as lesion detection, primarily due to their implicit property of inducing better calibration. Meanwhile, (with the exception of focal loss) both distribution and region/boundary-based loss functions perform equally well for anatomical or semantic segmentation tasks, such as quantification of organ shape, size and boundaries.
\end{abstract}

\section{Introduction}

\paragraph{Anatomical versus Diagnostic} Anatomical (organ-level) and diagnostic (pathology-level) segmentation are fundamentally different, with respect to their clinical objectives. For instance, within the clinical workflow of prostate MRI, segmentation of the prostate anatomy \{whole-gland (WG), transitional zone (TZ), peripheral zone (PZ)\} is used to estimate its volume, boundaries and geometry --enabling the calculation of prostate-specific antigen (PSA) density, guiding treatment planning and future interventions \cite{NEJM,PSAdPlanning,RadioTherap}. Meanwhile, segmentation of clinically significant prostate cancer (csPCa) is primarily used to detect the number of malignant lesions present in the prostate gland (if any), and characterize each instance for diagnosis, risk stratification and/or targeted biopsies (similar to PI-RADS guidelines \cite{PIRADSv2}). In other words, while the former anatomical task leans towards \textit{semantic} single-object quantification, the latter diagnostic task can be framed as \textit{panoptic} multi-object detection \cite{Panoptic,PanopticPCa}. Despite these differences, in medical image computing, these objectives are often treated and trained as near-identical segmentation tasks using U-Nets, and evaluated in a similar manner using the Dice Similarity Coefficient (DSC) \cite{DSC1,DSC2,DSC3,Jaeger}. Recent studies confirm that while DSC can be appropriate for single-object quantification, its inability to measure multi-object detection makes it unsuitable for lesion localization \cite{CommonPitfalls,DSC_LIM,DSCB,Panoptic}. We hypothesize that such considerations should not only be made during evaluation, but should also be incorporated at train-time by using task-specific loss function(s).

\vspace{-2mm}

\paragraph{Segmentation Loss Functions} We can distinguish between loss functions that minimize mismatch in underlying distributions and those that minimize mismatch in segmentation regions/boundaries \cite{LossOdyssey}. Distribution-based loss functions, such as the widely adopted cross-entropy (CE) loss, minimize dissimilarity between the target distribution (ground-truth) and its approximation (model predictions), over the course of training. Balanced cross-entropy (BCE) loss is an extension of CE, introducing class weights $\alpha$ to address imbalanced class frequencies in the dataset. Focal loss (FL) represents a further extension of BCE, introducing an additional hyperparameter $\gamma$ to differentiate between easy and difficult examples \cite{FocalLoss}. When $\gamma=0$, FL reduces down to CE or BCE depending on the value of $\alpha$. On the other hand, region or boundary-based loss functions are derived from integrals over segmentation regions (e.g. soft Dice loss \cite{VNet}) or distance metrics in the space of contours (e.g. boundary loss \cite{BoundaryLoss}). Unlike distribution-based losses, which can benefit from class re-weighting, the sum of Dice and boundary loss (DB) is implicitly impervious to class imbalance, and in turn, recommended for highly imbalanced datasets \cite{BoundaryLoss}.

\vspace{-1mm}

\section{Experiments and Analysis}

\vspace{-1.5mm}

\paragraph{Materials} We used 200 prostate bpMRI (T2W, high b-value DWI, ADC) exams from the publicly-available ProstateX dataset \cite{ProstateX}, paired with voxel-level delineations of WG, TZ, PZ and csPCa \cite{ProstateXAnnot}. All images were resampled to $0.5\times0.5\times3.0$ mm$^3$ resolution, center-cropped to $160\times160\times20$ voxels and intensity-normalized (T2W, DWI: \textit{z}-score; ADC: linear) \cite{E2E}, prior to usage. 

\vspace{-1mm}

\paragraph{Bayesian Segmentation Model} We used a probabilistic adaptation \cite{ProbUNet} of the deep attentive 3D U-Net, developed and validated specifically for prostate bpMRI in our previous work \cite{E2E}. Monte-Carlo dropout nodes were added to capture both \textit{epistemic} and \textit{aleatoric} uncertainty during inference (as recommended by \citet{MCProb}). Cosine annealing learning rate \cite{CosineAnnLRWR} (decaying from $10^{\textnormal{-}4}$ to $10^{\textnormal{-}7}$) and \textit{AMSGrad} optimizer \cite{AMSBound} were used to train the model. Data augmentations comprised of additive Gaussian noise (standard deviation $0$-$0.5$), horizontal flip, rotation ($\pm$$7.5$\textdegree), translation ($0$-$15$\% horizontal and/or vertical shifts) and scaling ($0$-$15$\%) centered along the axial plane. Fig. 1 illustrates the train-time schematic of the model, and its source code is made publicly-available at \textcolor{magenta}{\url{https://github.com/DIAGNijmegen/prostateMR_3D-CAD-csPCa}}.

\vspace{-1mm}

\paragraph{Experiments} We investigated four different loss policies (CE, BCE, FL, DB) as the segmentation loss function ($\textbf{\textit{L}}_\textbf{S}$) in our model for anatomical (\{WG, TZ, PZ\} in T2W MRI) and diagnostic (csPCa in bpMRI) segmentation. For the former task, $\alpha$ is set as [$0.05$, $0.30$, $0.65$] as per the empirical class distribution of the dataset, and for the latter task, $\alpha$ is set as [$0.75$, $0.25$] as per the findings of previous studies \cite{UCLA,E2E}. In both cases, $\gamma$ is set to its default value of $2.00$ as used in \cite{FocalLoss,UCLA,E2E}. All metrics were computed in 3D over 3 independent runs $\times$ 5-fold cross-validation $\times$ mean of 100 executions of probabilistic inference per image. Identical data splits were maintained for all configurations across both tasks.

\begin{figure}[b!]
\centering
\includegraphics[width=0.975\textwidth]{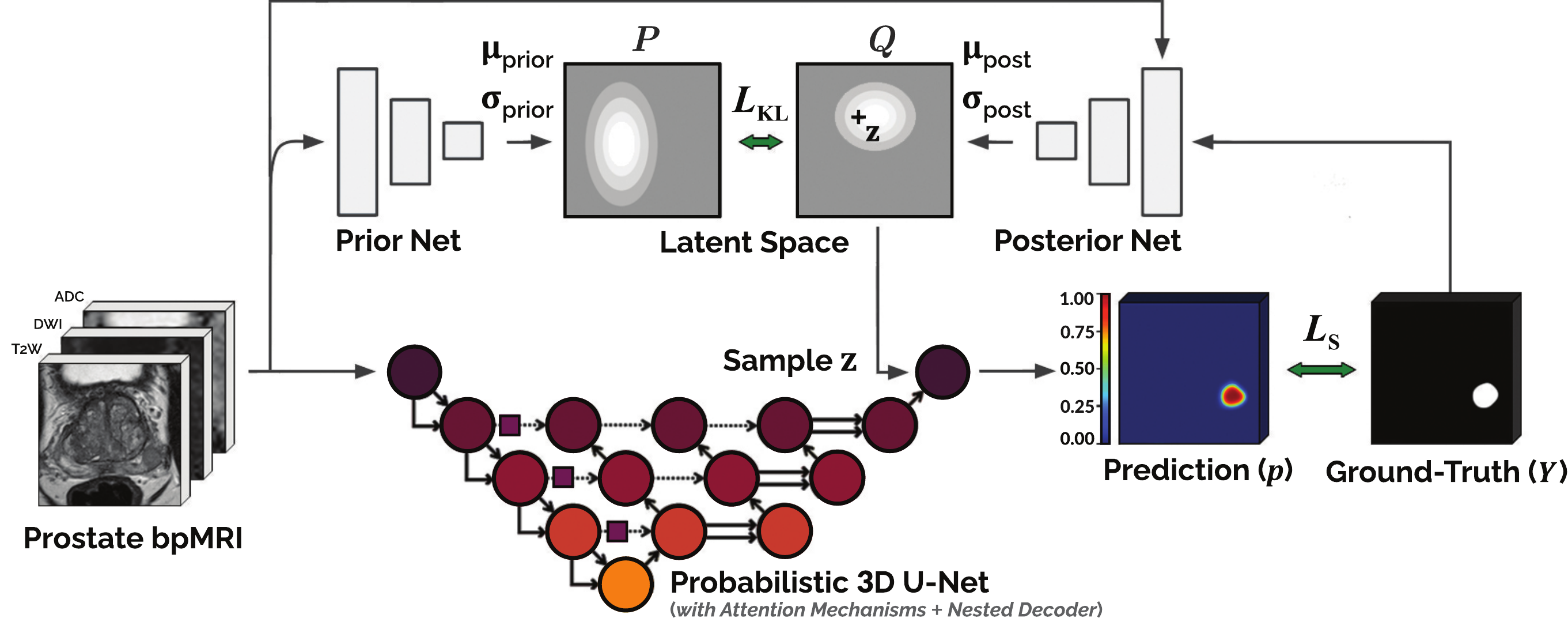}
\caption{Train-time schematic for the Bayesian segmentation model. $\textbf{\textit{L}}_\textbf{KL}$ denotes the Kullback--Leibler divergence loss between prior distribution $\textbf{\textit{P}}$ and posterior distribution $\textbf{\textit{Q}}$. $\textbf{\textit{L}}_\textbf{S}$ denotes the segmentation loss between prediction $\textbf{\textit{p}}$ and ground-truth $\textbf{\textit{Y}}$. For each execution of the model, one sample \textbf{z} $\in \textbf{\textit{Q}}$ (train-time) or \textbf{z} $\in \textbf{\textit{P}}$ (test-time) is drawn to predict one segmentation mask $\textbf{\textit{p}}$ \cite{ProbUNet}.}
\label{fig1}
\end{figure}

\newpage

\section{Results and Discussion}

\begin{figure}[h!]
  \centering
  \begin{tabular}{ccc}
  \centering
  \includegraphics[width=0.29\textwidth]{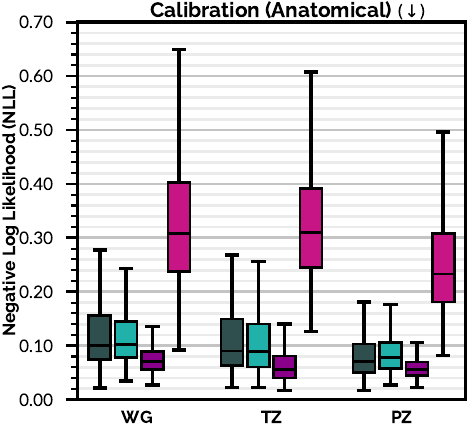} & \includegraphics[width=0.29\textwidth]{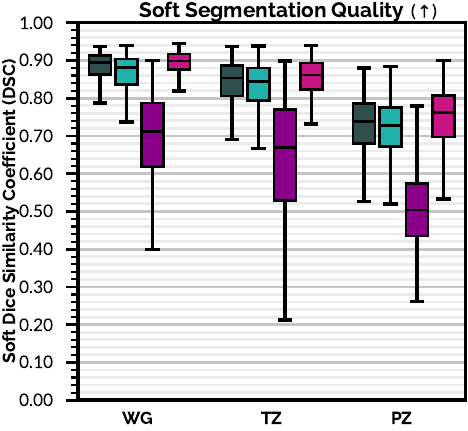} & \includegraphics[width=0.29\textwidth]{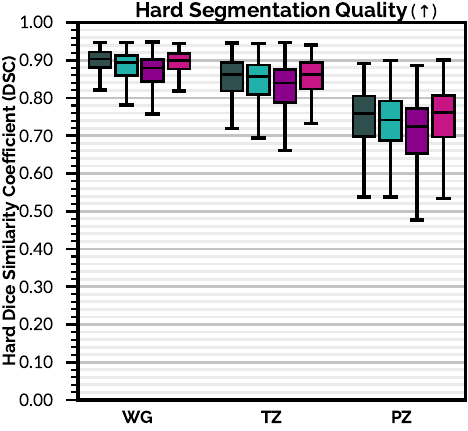} \\
  {\small \hspace{0.04\textwidth} \textbf{(a)}} & {\small \hspace{0.04\textwidth} \textbf{(c)}} & {\small \hspace{0.04\textwidth} \textbf{(d)}} \\
  \end{tabular}
  
  \vspace{1mm}
  
  \begin{tabular}{ccc}
  \centering
  \includegraphics[width=0.29\textwidth]{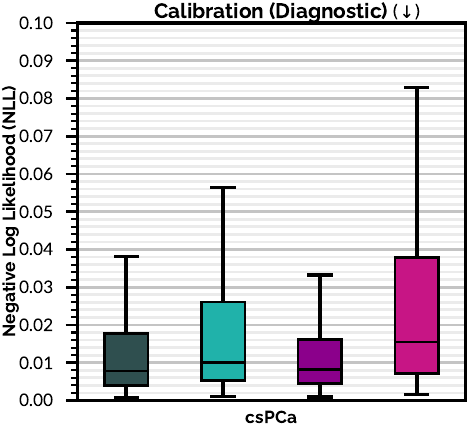} & \includegraphics[width=0.29\textwidth]{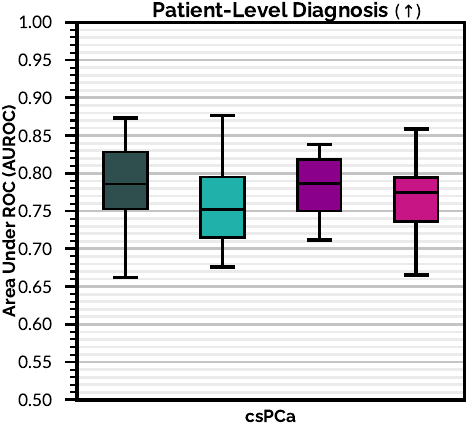} & \includegraphics[width=0.29\textwidth]{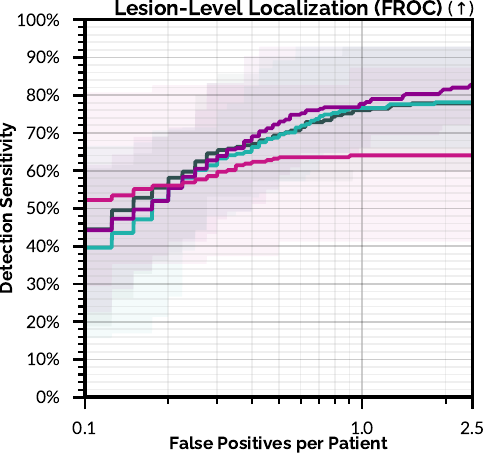} \\
  {\small \hspace{0.04\textwidth} \textbf{(b)}} & {\small \hspace{0.04\textwidth} \textbf{(e)}} & {\small \hspace{0.04\textwidth} \textbf{(f)}} \\
  \end{tabular}
  \centering
  \includegraphics[width=0.9\textwidth]{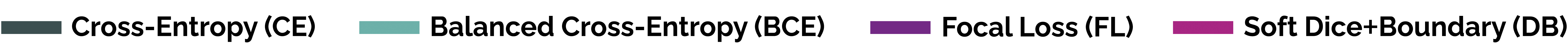}
  \caption{Anatomical \textbf{(top-row)} and diagnostic \textbf{(bottom-row)} segmentation of \{WG, TZ, PZ\} in T2W MRI and csPCa in bpMRI, respectively. For each $\textbf{\textit{L}}_\textbf{S}$ configuration, we evaluated $\textbf{(a, b)}$ model calibration with negative log-likelihood (NLL), $\textbf{(c, d)}$ segmentation quality with mean soft/hard Dice Similarity Coefficient (DSC) across all 600 observations (3 runs $\times$ 5 folds $\times$ 40 validation samples), $\textbf{(e)}$ patient-level diagnosis with Area Under Receiver Operating Characteristic (AUROC), $\textbf{(f)}$ lesion-level localization with Free-Response Receiver Operating Characteristic (FROC). Transparent areas in $\textbf{(f)}$ FROC indicate $95$\% confidence intervals. Arrows ($\uparrow$ or $\downarrow$) across $\textbf{(a-f)}$ indicate whether the corresponding metric should be maximized/minimized for ideal performance. }
  
  \vspace{-4mm}
\label{fig2}
\end{figure}

\paragraph{Calibration} We find that distribution-based loss functions (CE, BCE, FL) produce implicitly well calibrated predictions, in comparison to DB \cite{TMI_1} (as shown in Fig. 2 (a, b) and \textit{Appendix A}). Among them, FL exhibits notably better calibration. Our findings concur with that of \citet{FLeffects}, where the authors demonstrate that FL is formulated as such, that it minimizes dissimilarity between ground-truth and predicted distributions, while \textit{increasing entropy} of the latter --thereby limiting NLL overfitting and inducing both calibration and regularization. Such a property can also lead to more useful uncertainty estimates \cite{FLeffects,UnCa}.

\vspace{-3mm}
  
\paragraph{Anatomical Segmentation} From Fig. 2 (c), it is clear that while CE, BCE and DB achieve similar segmentation quality with respect to raw \textit{softmax} predictions (\{WG: $81$-$89$; TZ: $82$-$85$; PZ: $71$-$74$\} mean DSC), FL performs worse (\{WG: $70$; TZ: $64$; PZ: $50$\} mean DSC). When considering binarized \textit{argmax}(\textit{softmax}) predictions, as seen in Fig. 2 (d), overall FL performance is only marginally lower than that of CE, BCE and DB (\{WG: $88$-$90$; TZ: $84$-$86$; PZ: $72$-$76$\} mean DSC). We attribute this decline in FL performance to its high predictive uncertainty near class boundaries (as seen in \textit{Appendix A}), which limits its ability to produce precise contour definitions.

% pAUC improvement: (1.740573251706163 + 1.713157412770611 + 1.8363116623299842)/3 - 1.5111661575563213 = 0.2521812847125981
% max sens improvement: (0.7781 + 0.7811 + 0.8319)/3 - 0.6404  = 0.1566333333

\vspace{-3mm}

\paragraph{Diagnostic Segmentation} From Fig. 2 (e, f), we observe that while all loss functions are similar in patient-level diagnosis ($76$-$78$ mean AUROC), results indicate that distribution-based loss functions, especially FL, perform substantially better than region/boundary-based losses (DB) at lesion detection (with an average increase of $15.7\%.$ in maximum detection sensitivity and $0.25$ in partial area under FROC between $0.1$-$2.5$ false positives per patient). We attribute this improvement to the implicit train-time calibration induced by FL, which can facilitate better risk stratification and higher detection sensitivities across difficult diagnostic tasks (e.g. via retaining low-confidence lesion predictions, as opposed to muting them out --refer to \textit{Appendix A}). We believe that the growing success of csPCa detection models trained using FL derivatives (such as the \textit{FocalNet} \cite{UCLA}, among others \cite{E2E,ISBI_Cao,ProCDet}) can, in part, be attributed to this phenomenon rather than architectural enhancements or class weighting alone (note BCE performs near-identical to CE, if not worse, across both tasks). On the other hand, while DB loss is a natural fit for maximizing DSC, it is ill-posed to optimize panoptic segmentation objectives \cite{Panoptic,CommonPitfalls}. Subsequently, its miscalibration (refer to Fig. 2 (a, b)) or polarized predictions (refer to \textit{Appendix A}), translate to a relatively flat FROC curve with lower maximum detection sensitivity (refer to Fig. 2 (f)) --presumably detection of low-confidence, difficult or small lesions are skipped in favour of maximizing confidence and overlap of clear lesions \cite{CommonPitfalls}. 

In conclusion, we recommend distribution-based loss functions (in particular, FL) for diagnostic or panoptic segmentation tasks. For anatomical or semantic segmentation tasks, we observe that although distribution-based losses improve calibration, (with the exception of FL) they are equivalent to region/boundary-based losses in terms of standard/hard DSC. Hence, CE, BCE, DB or presumably their composites (as applied by the \textit{nnU-Net} \cite{nnUNet} and investigated by J. Ma et al. \cite{LossOdyssey}), can be recommended. Further analyses are required, using larger datasets and multiple medical imaging modalities, to draw out definitive conclusions.

\vspace{-2mm}

\section*{Broader Impact}

\vspace{-2mm}

Prostate cancer is one of the most prevalent cancers in men worldwide \cite{PCaStat2019}. In the absence of experienced radiologists, its multifocality, morphological heterogeneity and strong resemblance to numerous non-malignant conditions in MR imaging, can lead to low inter-reader agreement ($<50\%$) and sub-optimal interpretation \cite{mpMRI,LimitedPIRADS2,LimitedPIRADS3}. The development of automated, reliable detection algorithms has therefore become an important research focus in medical image computing ---offering the potential to support radiologists with consistent quantitative analysis, improve diagnostic accuracy, and in turn, minimize unnecessary biopsies in patients \cite{UnnBiop,RadPRvsDL}. 

To the best of our knowledge, this study has no foreseeable negative societal effects.

\vspace{-2mm}

\begin{ack}

\vspace{-2mm}

This research is supported, in parts, by the European Union H2020: ProCAncer-I project (EU grant 952159), Health{\textasciitilde}Holland (LSHM20103), ContextVision AB (Linköping, Sweden) and Siemens Healthineers (CID: C00225450).
\end{ack}

%\section{Acknowledgement and Disclosure of Funding}
%ProCAncer-I, Siemens Healthineers

{
\fontsize{8.75pt}{8.75pt}\selectfont
\bibliographystyle{unsrtnat}
\bibliography{main.bib}
}

\newpage

%%%%%%%%%%%%%%%%%%%%%%%%%%%%%%%%%%%%%%%%%%%%%%%%%%%%%%%%%%%%

\appendix

\section{Appendix: Model Predictions}

\vspace{-4mm}

\begin{figure}[h!]
\centering
\includegraphics[width=\textwidth]{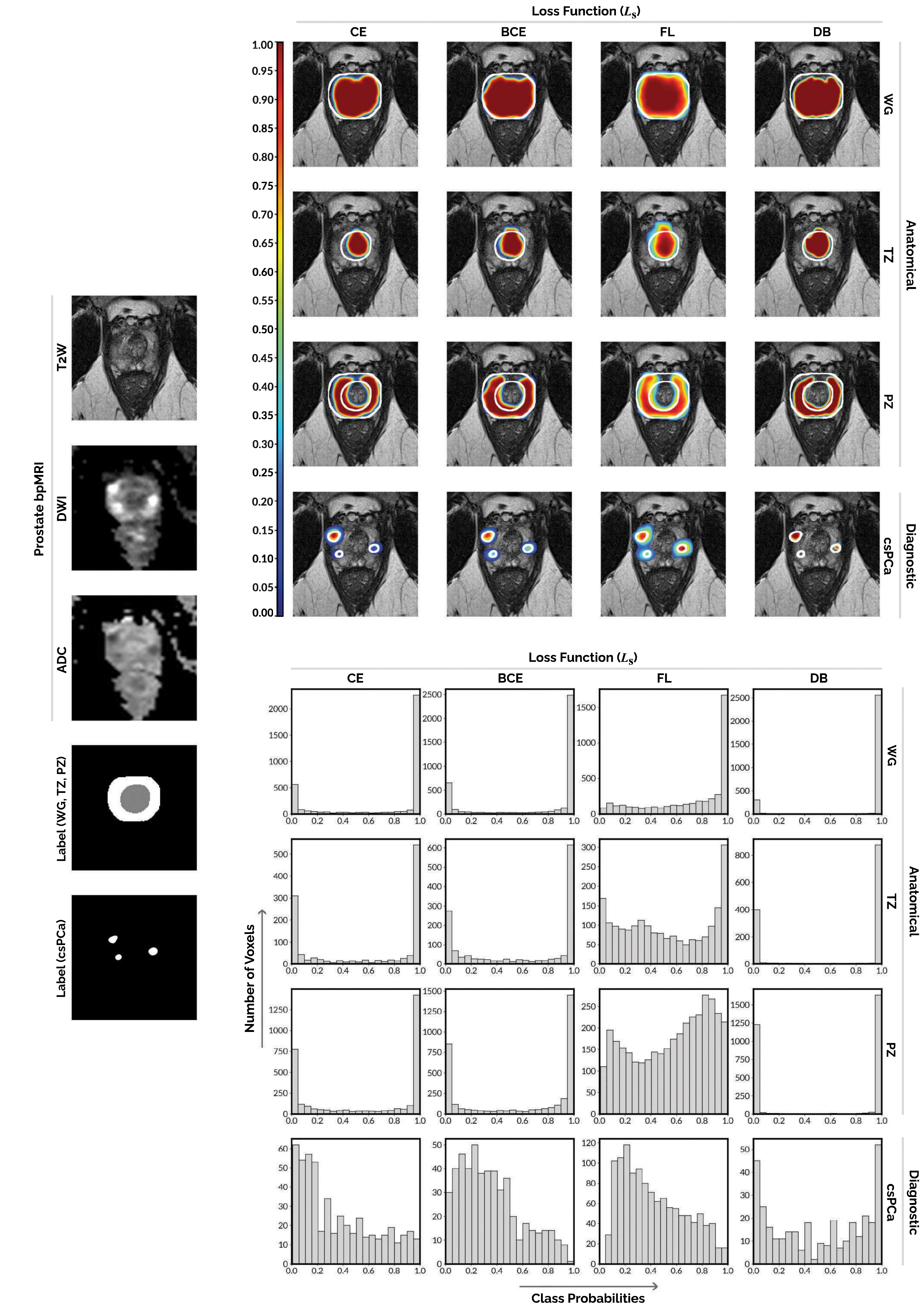}
\caption{Anatomical (WG, TZ, PZ) and diagnostic (csPCa) segmentations for a single validation scan, across all four loss functions (CE, BCE, FL, DB), are shown above. Predictions are overlaid on the T2W scan as reference, where white lines indicate the corresponding ground-truth annotation for the given task. For each case, a histogram of class probabilities over the predicted segment(s) has been illustrated below, indicating the degree of implicit calibration induced by the loss function.}
\label{fig3}
\end{figure}

\end{document}